\documentstyle[a4,psfig]{article}

\begin{document}
\title{R-mode oscillations of differentially and rapidly rotating Newtonian 
polytropic stars}
\author{Shigeyuki Karino$^1$ \footnote{E-mail: karino@valis.c.u-tokyo.ac.jp},
        Shin'ichirou Yoshida$^2$ , Yoshiharu Eriguchi$^1$ \\
	\small $ ^1$ Department of Earth Science and Astronomy,\\
        \small Graduate School of Arts and Sciences,\\
        \small University of Tokyo, Komaba, Meguro, Tokyo 153-8902, Japan\\
	\small $ ^2$ SISSA, Via Beirut 2-4, 34013 Trieste, Italy
}

\date{\today}
\maketitle

\begin{abstract}
For the analysis of the r-mode oscillation of hot young neutron stars, it is
necessary to consider the effect of {\it differential rotation}, because 
viscosity is not strong enough for differentially rotating young neutron stars 
to be lead to uniformly rotating configurations on a very short time scale 
after their birth. In this paper, we have developed a numerical scheme to 
solve r-mode oscillations of differentially rotating polytropic inviscid stars. This is 
the extended version of the method which was applied to compute r-mode 
oscillations of uniformly rotating Newtonian polytropic stars. By using this 
new method, we have succeeded in obtaining eigenvalues and eigenfunctions of 
r-mode oscillations of differentially rotating polytropic stars.  

Our numerical results show that as the degree of differential rotation is 
increased, it becomes more difficult to solve r-mode oscillations for slightly 
deformed configurations from sphere compared to solving r-mode oscillations 
of considerably deformed stars. One reason for it seems that for slightly 
deformed stars corotation points appear near the surface region if the degree 
of differential rotation is strong enough. This is similar to the 
situation that
the perturbational approach of r-mode oscillations for {\it slowly rotating} 
stars in {\it general relativity} results in a singular eigenvalue problem.

\end{abstract}


\section{Introduction}

Recent intensive as well as extensive investigations of the r-mode instability 
have revealed its important role in neutron star physics (\cite{a98,fm98};
for a review see e.g. \cite{fl99,ak00}; for r-mode oscillations see 
\cite{pp78,pbr81,s82}).  At the early stage of those investigations, the 
r-mode oscillations were found to become significantly unstable for some range 
of the core temperature \cite{lom98,olcsva98,aks99}.  However, models used in 
those papers were too simplified to deduce a decisive conclusion for real 
neutron stars because one could only obtain r-mode oscillations of 
{\it uniformly} and {\it slowly} rotating Newtonian polytropic stars and 
apply the results to realistic neutron stars.

Actual neutron stars are more complicated configurations.
There are many factors to be included to get a definite answer about the
effect of the r-mode instability on them. For example, neutron stars are 
general relativistic objects, they are generally expected to be rapidly 
rotating at their birth, the physical states of the interior cannot be 
quantitatively approximated by polytropic relations, the surface of the 
neutron stars is not considered to be fluid, they have magnetic fields
and so on.

The rotation law is one of such factors.  Although in almost all papers about 
the r-mode instabilities neutron stars are assumed to rotate uniformly, 
newly born neutron stars whose ages are less than 1 year should be 
considered to be rotating differentially. 
Some neutron stars are newly formed from core collapses of massive stars on a 
dynamical time scale without enough time to redistribute the angular momentum 
due to viscosity and rotate differentially.  
(At some stages of recycled neutron stars, they may begin to rotate 
differentially due to instability of f-mode oscillations via gravitational 
wave emission because angular momentum increase due to accretion onto old 
neutron stars leads them to bifurcation points to configurations within which 
there exists internal motion. However, this instability results in 
non-axisymmetric configurations and they are out of the scope of the present 
paper. )
Thus we need to study the effect of differential rotation of newly born 
neutron stars on the r-mode oscillations. 

Moreover, some authors discuss the onset of differential rotation of a
star due to self-induction by r-mode oscillations. Spruit \cite{spr99} argued 
that the back-reaction of gravitational radiation by the r-mode oscillation 
on the stellar fluid may induce the differential rotation of the star. 
Rezzolla et al. \cite{rls00,rlms00a,rlms00b} studied a drift induced by the r-mode oscillation 
of the stellar fluid which is initially uniformly rotating, and concluded 
that the r-mode instability may lead to the differential rotation of the star.
By using numerical simulation of stellar hydrodynamics, Lindblom 
et al. \cite{ltv00} studied non-linear evolution of an initially uniformly 
rotating star by the r-mode instability and found that the stellar fluid 
develops strong differential rotation. Thus in the general context of the 
r-mode instability, it seems to be important to investigate the r-mode 
oscillations of differentially rotating stars.

In our present investigation, by extending the numerical method which we 
developed to compute r-mode oscillations of uniformly and rapidly rotating 
polytropes \cite{ykye00,kyye00}, we have succeeded in obtaining sequences 
of r-mode oscillations for differentially rotating polytropic stars 
(in a different context, the r-mode oscillations of configurations 
with {\it slightly} differential rotation are treated in \cite{w98,w00}).
From our numerical results for differentially rotating stars, we have found 
that, if the degree of differential rotation is small, r-mode oscillations of 
the ordinary type can be obtained. Here r-modes of the ordinary type mean
that the eigenfrequency has discrete spectrum.  However, 
when the degree of differential rotation becomes sufficiently large, 
no r-mode oscillations 
of the ordinary type seem to be allowed, because in the surface region of 
the star corotation points appear and the basic equation becomes singular 
there.  

Since the purpose of this paper is to show the characteristic feature 
of the r-mode oscillation of differentially rotating stars, we only present 
a limited results of numerical computations, i.e. for the $m = 2$ mode, 
where $m$ is the azimuthal mode number of the perturbation.

\section{Formulation of the problem}

We study the effect of differential rotation on the r-mode oscillations 
by solving the linearized equations of fluid motions of stars.
The scheme employed in this paper is the extended version of that developed
to analyze the r-mode oscillation of uniformly rotating Newtonian
polytropes \cite{ykye00,kyye00}. Therefore, we will explain the scheme
only briefly.

\subsection{Assumptions on unperturbed stars}

We construct axisymmetric equilibrium configurations of differentially 
rotating polytropic inviscid stars in Newtonian gravity by employing the 
Straight-Forward-Newton-Raphson method \cite{em85}. 
Here the polytropic relation is expressed by the 
following formula:
\begin{equation}
p = K \rho^{1+\frac{1}{N}} ,
\end{equation}
where $p$, $\rho$, $K$ and $N$ are the pressure, the density, the polytropic 
constant and the polytropic index, respectively.  
In this paper polytropes with $N = 0.5, 1.0$ and $1.5$ are investigated.

As for the rotation law of differentially rotating unperturbed stars,
two kinds of angular velocity, $\Omega$, distribution are employed:

\begin{equation}
\Omega = \frac{\Omega_{\rm c} A^2}{(R/R_{\rm eq})^2 + A^2}, 
\label{eq:rot}
\end{equation}
and 
\begin{equation}
\Omega = \frac{\Omega_{\rm c} A}{\sqrt{(R/R_{\rm eq})^2 + A^2}}.
\label{eq:rot2}
\end{equation}
where $R$ is the distance from the rotation axis, $R_{\rm eq}$ is the
equatorial radius of the star and $\Omega_{\rm c}$ is the central angular
velocity.  The quantity $A$ is a parameter which represents the degree of 
differential rotation. The rotation becomes more differential as $A$ becomes 
smaller. On the other hand, when we consider the limit of $A \to \infty$, the 
stellar rotation tends to uniform rotation in both cases. 
For the region where $R/R_{\rm eq} \gg A$, the former law, Eq.~(\ref{eq:rot}), 
tends to that of constant specific angular momentum distribution, while the 
latter, Eq.~(\ref{eq:rot2}), tends to that of constant linear velocity 
distribution. Thus we will call the former as a j-constant rotation law and 
the latter as a v-constant rotation law, hereafter.

\subsection{Linearized basic equations for perturbed stars}

As mentioned before, the scheme to handle the r-mode oscillation of 
differentially rotating polytropes is the extended version of that to solve 
the perturbed fluid equations for uniformly rotating polytropes by the 
Newton-Raphson iteration scheme \cite{ykye00,kyye00}. 

Our basic equations consist of linearized parts of the following equations:
1) the continuity equation, 
2) the $\theta$-component of the vorticity conservation equation, 
   or the compatibility equation between the $r$- and $\varphi$-components
   of the equation of motion,
3) the $\varphi$-component of the vorticity conservation equation, 
   or the compatibility equation between the $r$- and $\theta$-components
   of the equation of motion,
and 
4) the $\varphi$-component of the equation of motion.
The linearized gravitational potential in the integral form is used in
the $\varphi$-component of the equation of motion. 
Here, the spherical coordinates $(r, \theta, \varphi)$ are used.

Concerning the perturbed quantities, we assume the harmonic expansion of the 
physical quantities as
\begin{equation}
\delta f (r,\theta,\varphi,t)
= \sum_m \exp(-i\sigma t + i m \varphi) f_m(r,\theta),
\end{equation}
where $f$ denotes a certain physical quantity, $\delta$ is the Euler
perturbation of the corresponding quantity, $\sigma$ is the frequency of 
the oscillation and $m$ is the mode number.  
Because of the stationary and axisymmetric nature of unperturbed stars 
the equations are separable
with respect to the variables $t$ and $\varphi$. Thus our basic equations 
are characterized by $\sigma$ and $m$. The perturbed 
fluid equations are expressed as follows:

\begin{equation}
\rho_0 \frac{\partial \delta u_r}{\partial r}
+ \Bigl( \frac{2 \rho_0}{r} + \frac{\partial \rho_0}{\partial r} \Bigr)
\delta u_r
+ \frac{\rho_0}{r} \frac{\partial \delta v_{\theta}}{\partial \theta}
+ \frac{1}{r} \Bigl( \rho_0 \cot \theta
+ \frac{\partial \rho_0}{\partial \theta} \Bigr) \delta v_{\theta}
+ \frac{m \rho_0}{r \sin \theta} \delta w_{\varphi}
= ( \sigma - m \Omega ) \delta \rho ,
\end{equation}

\begin{eqnarray}
\frac{r \sin^2 \theta}{m} \Bigl( 2 \Omega
+ r \frac{\partial \Omega}{\partial r} \Bigr)
\frac{\partial \delta u_r}{\partial r} 
+ \frac{1}{m} \Bigl[ m ( m \Omega - \sigma ) + \sin^2 \theta \Bigl( 2 \Omega
+ 4 r \frac{\partial \Omega}{\partial r}
+ r^2 \frac{\partial^2 \Omega}{\partial r^2} \Bigr) \Bigr] \delta u_r
\nonumber \\
+ \frac{r \sin^2 \theta}{m} \Bigl( 2 \Omega \cot \theta
+ \frac{\partial \Omega}{\partial \theta} \Bigr)
\frac{\partial \delta v_{\theta}}{\partial r}
+ \frac{r \sin^2 \theta}{m} \Bigl[ 2 \Omega \cot \theta
+ \frac{\partial \Omega}{\partial \theta}
+ 2 r \cot \theta \frac{\partial \Omega}{\partial r}
+ r \frac{\partial^2 \Omega}{\partial r \partial \theta} \Bigr]
\delta v_{\theta}
\nonumber \\
+ \frac{r \sin \theta}{m} ( m \Omega - \sigma )
\frac{\partial \delta w_{\varphi}}{\partial r}
+ \frac{\sin \theta}{m} \Bigl[ ( m \Omega - \sigma ) + 2 m \Omega
+ m r \frac{\partial \Omega}{\partial r} \Bigr] \delta w_{\varphi}
= 0 ,
\end{eqnarray}

\begin{eqnarray}
(\sigma - m \Omega) \frac{\partial \delta u_r}{\partial \theta}
- m \frac{\partial \Omega}{\partial \theta}
+ r (m \Omega - \sigma) \frac{\partial \delta v_{\theta}}{\partial r}
+ \Bigl[ (m \Omega - \sigma) + m r \frac{\partial \Omega}{\partial r} \Bigr]
\delta v_{\theta}
\nonumber \\
+ 2 r \cos \theta \Omega \frac{\partial \delta w_{\varphi}}{\partial r}
- 2 \sin \theta \Omega \frac{\partial \delta w_{\varphi}}{\partial \theta}
+ 2 \Bigl( r \cos \theta \frac{\partial \Omega}{\partial r}
- \sin \theta \frac{\partial \Omega}{\partial \theta} \Bigr) \delta w_{\varphi}
= 0,
\end{eqnarray}

\begin{eqnarray}
K \Bigl( 1 + \frac{1}{N} \Bigr) \rho_0^{\frac{1}{N} - 1}
\frac{m}{r \sin \theta} \delta \rho + \frac{m}{r \sin \theta} \delta \phi
+ \sin \theta \Bigl( 2 \Omega + r \frac{\partial \Omega}{\partial r} \Bigr)
\delta u_r
\nonumber \\
+ \Bigl( 2 \Omega \cos \theta + \sin \theta 
\frac{\partial \Omega}{\partial \theta} \Bigr) \delta v_{\theta}
+ (m \Omega - \sigma) \delta w_{\varphi} = 0.
\end{eqnarray}
\begin{eqnarray}
\delta \phi=-4\pi G \sum_{n,m} \int_0^{\frac{\pi}{2}}
d\theta ' \sin \theta ' \frac{(n-m)!}{(n+m)!}
P^m_n (\cos \theta) P^m_n (\cos \theta ') \nonumber \\
\times \int_0^{r_s(\theta ')}d r 'r ^{\prime 2} f_n (r,r ')
\delta\rho \nonumber \\
-4\pi G \sum_{n,m} \int_0^{\frac{\pi}{2}}
d\theta ' \sin \theta ' \frac{(n-m)!}{(n+m)!}
P^m_n (\cos \theta) P^m_n (\cos \theta ') \nonumber \\
\times f_n (r,r_s(\theta ')) \rho_0(r_s(\theta '),\theta ')
\delta r_s (\theta ') \ ,
\label{potential}
\end{eqnarray}
where $r_s (\theta)$ is the surface radius of the equilibrium configuration
and $\delta r_s$ is the change of the surface. Here, functions $f_n(r,r')$
are defined as

\begin{equation}
f_n (r,r') = \left\{ 
\begin{array}{ll}
\frac{1}{r}\left(\frac{r'}{r}\right)^n & (r'< r)\\
\frac{1}{r'}\left(\frac{r}{r'}\right)^n & (r'\geq r),
\end{array}
\right.
\end{equation}

and $\rho_0$ denotes the mass density of the equilibrium 
states, whereas $\delta u_r$, $\delta v_{\theta}$, $\delta w_{\varphi}$,
$\delta \rho$ and $\delta \phi$ are the Euler perturbations of the $r$-, 
$\theta$- and $\varphi$-components of the velocity in the corresponding 
orthonormal frame, the density and the gravitational potential, respectively.

In these equations, we have made use of the following adiabatic perturbation
between the pressure and the density:
\begin{equation}
\frac{\delta p}{p} = \left(1 + \frac{1}{N}\right) \frac{\delta \rho}{\rho},
\end{equation}
where $\delta p$ is the Euler perturbation of the pressure.

\subsection{Boundary conditions}

Since we treat infinitesimal oscillations around equilibrium configurations,
all the perturbed quantities must behave regularly throughout the whole space.
In particular, the Eulerian change of the density and the Eulerian change of the
gravitational potential must be regular functions of the position. 

As for the perturbed flow velocity, the boundary condition on the surface
can be expressed as
\begin{equation}
\frac{\partial \rho_0}{\partial r} \Big( \delta u_r
- \frac{1}{r_s} \frac{d r_s}{d \theta} \delta v_{\theta} \Bigr)
+ (m \Omega - \sigma) \delta \rho = 0 .
\end{equation}
%
This condition comes from a requirement that the fluid element 
on the equilibrium surface
is displaced to the perturbed surface.

\section{Numerical results}

We can obtain eigenvalues and eigenfunctions by solving numerically the basic 
equations mentioned above by the Newton-Raphson iteration scheme, together 
with the boundary condition. In actual computations, we use $44$ mesh points
in $r$-direction and $11$ mesh points in $\theta$-direction.

In Figure~1, the eigenvalues of equilibrium sequences with different values
of the parameter $A$ are plotted against the 
ratio of the rotational energy, $T$, to the absolute value of the 
gravitational energy, $W$, for the $m = 2$ mode of $N = 1$ polytropes. 
In this figure, for the sake of simplicity, the eigenvalues are normalized 
by using the central angular velocity.  This choice of the normalization 
constant results in decreasing of the eigenfrequencies for the smaller 
value of $A$ or for the larger degree of differential rotation.  
It is clearly seen that the eigenvalues decrease as the ratio $T/|W|$ 
increases for a given value of $A$. 
Terminal points of sequences at the larger end of $T/|W|$ correspond
to final models beyond which our present 
code could not give converged solutions.

Figure~2 shows $\theta$-component of the perturbed  
velocity on the equatorial plane for the $m=2$ oscillation of $N = 1$ 
polytropes with different degrees of nonuniform rotation.  The horizontal 
axis is the normalized stellar radius. The equilibrium stars are rotating 
rather rapidly, 
i.e. $r_{\rm p} = 0.7$. Here $r_{\rm p}$ is the axis ratio which is defined by
\begin{equation}
r_{\rm p} = r_{\rm ax}/r_{\rm eq} , 
\end{equation}
where $r_{\rm ax}$ and $r_{\rm eq}$ are the stellar radius along 
the rotational axis and that on the equatorial plane, respectively.

Three curves in Figure~2 correspond to equilibrium models of rigid rotation, 
differential rotation with the parameter $A = 1.3$ and extremely differential 
rotation with $A = 0.6$, respectively. 
The values of $T / |W|$ are almost the same for these models,
i.e. $ T/|W| = 0.073, 0.082$ and $0.080$ for the uniformly rotating
model, the model with $A = 1.3$  and the model with $A = 0.6$, respectively.
Thus we can regard these three models as those with the same rotational
amount but different degrees of differential rotation.
From this figure, we can see that if we take the differential 
rotation into consideration, oscillatory motions of the stellar fluid are 
confined only in the narrow layer near the surface of the star for a model 
with larger degree of differential rotation. In other words, for 
differentially rotating models, if the degree of differential rotation is
large, only the fluid near the surface region can oscillate appreciably but 
the inner bulk of the star almost remains at its original position.
This behavior can be seen not only in the equatorial plane but also 
in all $\theta-$ directions. 

In Figure~\ref{fig:EOS}, eigenfrequencies of r-mode oscillations for the 
$m = 2$ mode of differentially rotating polytropes with several polytropic 
indices are plotted against the value of $T/|W|$, i.e. $N = 0.5, 1.0$ and 
$1.5$.  Here we have used the rotation law (\ref{eq:rot}) with $A = 1.3$.  
Terminal points of theses sequences are final models beyond which our code 
could not give solutions to the linearized basic equations because of very 
slow convergence.  From this figure, it is clear that, although the values of 
$\sigma/\Omega_{\rm c}$ are not the same, the tendency of decrease of
the normalized eigenfrequency as increase of $T/|W|$ is the same.

For oscillations of differentially rotating stars, we need to pay attention 
to emergence of corotation points at some places in the stellar interior 
because appearance of corotation points corresponds to a singular behavior 
of the basic equations (see e.g. \cite{il90}).  Here corotation points are 
defined as points at positions where the following condition is satisfied:
\begin{equation}
{\hat \sigma} \equiv \sigma - m \Omega = 0 .
\end{equation}

The condition of appearance of corotation points can be expressed by a simple 
inequality for differentially rotating stars. For the j-constant rotation 
law, Eq.~(\ref{eq:rot}), the quantity $\hat \sigma$ is rewritten as
\begin{equation}
{\hat \sigma}
= \Omega_{\mathrm{c}} \Bigl( \frac{\sigma}{\Omega_{\mathrm{c}}}
- \frac{m A^2}{(R/R_{\mathrm{eq}})^2 + A^2} \Bigr) \ . \label{eq:cond1}
\end{equation}
If this quantity $\hat \sigma$ becomes zero within the star, we cannot solve 
the eigenvalue problem because the matrix which governs the problem becomes
singular as mentioned above. From Eq.~(\ref{eq:cond1}), the quantity
$\hat \sigma$ takes its minimum value on the rotational axis 
and its maximum value on the equatorial surface. Hence,
\begin{equation}
\Omega_{\mathrm{c}} \Bigl( \frac{\sigma}{\Omega_{\mathrm{c}}} - m \Bigr)
\leq \sigma - m \Omega \leq 
\Omega_{\mathrm{c}} \Bigl( \frac{\sigma}{\Omega_{\mathrm{c}}}
- \frac{m A^2}{1 + A^2} \Bigr).
\end{equation}
For $m = 2$ r-modes of $N = 1$ polytropes, as seen from Fig.~1, the lower limit
is always negative in the star. Thus the condition for which the
quantity $\hat \sigma$ becomes zero can be obtained by requiring
that the upper limit of $\hat \sigma$ must be positive inside the star.
As a result, the condition that corotation points exist in the star can be
written as
\begin{equation}
\frac{\sigma}{\Omega_{\mathrm{c}}} \geq \frac{2 A^2}{1 + A^2}, 
\end{equation}
where $m = 2$ is substituted because we only discuss $m = 2$ oscillation
modes in this paper. In order to find out whether the corotation points 
appear in the star or not, we need to check this condition for oscillations
of differentially rotating stars with the j-constant rotation law.

In Figure~4, values of eigenfrequencies which satisfy the {\it equality} of 
the above condition are plotted against the value of $\log A$.  We will call
this curve as a critical curve. If the eigenfrequency of a certain star is
located above the critical curve, a corotation point appears in the star.
On the other hand, if the eigenfrequency is below the critical curve, 
no corotation point appears. In this figure, eigenfrequencies of two 
equilibrium sequences, sequences with $r_{\rm p} = 0.95$ and with 
$r_{\rm p} = 0.70$, are also plotted.  
As seen from this figure, the eigenvalues locate well below the critical 
curve when the degree of differential rotation is small. When we decrease 
the value of $A$, i.e. increase the degree of differential rotation, 
however, the eigenfrequencies approach to the critical curve. We can infer 
from the result here that, 
for extremely slowly rotating stars for which the standard 
slow-rotation approximation is applied, the permitted range of the parameter 
$A$ for the existence of discrete modes is {\it narrower} than that for rapidly 
rotating stars. 

We show the ratio of the eigenfrequency to the critical value for appearance 
of corotation points in Figure~5. From this figure, it is clear that 
eigenfrequencies are approaching toward the critical curve monotonically as 
the value of $A$ is decreased. For nearly spherical configurations, i.e. 
the sequence with $r_{\rm p}=0.95$, the eigenfrequency of the model with 
$A=0.615$ reaches up to 97\% of the critical value.  On the other hand, 
for rapidly rotating models ($r_p = 0.70$), eigenfrequency reaches only 93\% 
of its critical value, when $A=0.585$.

For slowly rotating stars ($r_{\rm p}\sim 1$), sooner or later, the eigenvalues
seem to cross the critical curve, though it is very difficult to get converged 
solutions of oscillations by using our numerical code in such cases.
This is not related to the ability of the code because r-mode oscillations
can be easily obtained for nearly spherical configurations with larger 
values of $A$.

For the v-constant rotation law, the condition that corotation points 
of $m = 2$ mode appears in the star is written as
\begin{equation}
\frac{\sigma}{\Omega_{\mathrm{c}}} \geq \frac{2 A}{\sqrt{1 + A^2}}.
\end{equation}

In Fig.~6, values of eigenfrequencies which satisfy the {\it equality} of 
the above condition are plotted against the value of $\log A$ as in Fig.~4
of the j-constant rotation law. The curves denoted by rapid rotation
and slow rotation correspond to the eigenfrequencies of the equilibrium 
sequences with $ r_{\rm p} = 0.6 $, and $0.95$, respectively.

As seen from this figure, the eigenfrequency reaches only up to about 71 \% 
of the critical frequency even for the model with the same axis ration
of $r_{\rm p} = 0.95$, 
$ A = 0.25 $ and $T/|W| = 0.010$. It should be noted that this value of 
$T/|W| = 0.010$ is {\it smaller} 
than the value of $T/|W| = 0.031$ for the model 
with $r_{\rm p} = 0.95$ and $A = 0.65$ for the j-constant rotation law. It
implies that corotation points are more likely to appear for the rotation 
law with a steeper angular velocity distribution from the rotational axis to 
the equatorial surface.

\section{Discussion and summary}

\subsection{Influence of differential rotation on the spin evolution
of young hot neutron stars}
It is important to know how far  differential rotation will change the spin
evolution scenario of young hot neutron stars by the r-mode instability.  
In order to get precise answer to this problem, we need to perform time evolutionary
computations of rotating stars.  However, it is a difficult task to do even
for uniformly rotating stars.  For the estimation of spin evolution of 
uniformly rotating neutron stars, therefore, only the growth rates of the 
unstable modes have been investigated by evaluating the time scales of
r-mode instability due to gravitational emission and 
damping times of the modes due to viscosity (see e.g. \cite{lom98,olcsva98}).

Although, for uniformly rotating stars, the scheme to evaluate time scales 
has been already established by using the energy of the mode as measured in 
the rotating frame \cite{fs78,il91}, which is defined as a quadratic form 
of Eulerian perturbations, there is no corresponding definition of the energy 
functional which can be used to determine the stability of the oscillation 
modes for {\it differentially rotating} stars. Thus in this paper we will 
discuss only a tendency of evolution of differentially rotating stars via
instability of the r-mode.

There are two main reasons why the r-mode instability for the uniformly 
rotating stars becomes important for spin evolution of young hot neutron 
stars.  First, the growth rate of the instability due to gravitational wave 
emission is so large that the spin change of neutron stars would occur on a 
very short time scale. Second, due to the characteristic feature of r-mode 
oscillations, the stabilizing effect due to (bulk) viscosity is much
 weaker than 
that for the f-mode oscillations.  Although it is not easy to estimate 
quantitatively the effect of these two mechanisms for differentially rotating 
stars,  we note that the time scale of gravitational emission from
differentially rotating stars may be larger compared with that for
corresponding uniformly rotating stars.

The energy loss rate due to gravitational wave emission can be written as
\begin{equation}
\frac{dE}{dt} = - \frac{1}{32 \pi} \sum_{l \ge 2, l\ge m} \left(
\left| ^{(l+1)}D_{lm}\right|^2 + \left| ^{(l+1)}J_{lm}\right|^2\right) \ , 
\end{equation}
where $D_{lm}$ and $J_{lm}$ are the mass multipole moments and the current 
multipole moments, respectively \cite{t80}.
Here $^{(l+1)}D_{lm}$ and $^{(l+1)}J_{lm}$
are the $(\ell+1)$-th time derivatives of the corresponding quantities.
These two multipole moments are defined as:
\begin{eqnarray}
D_{lm} & \equiv & \int r^l \rho Y_l^{*m} d^3x \ , \\ 
J_{lm} & \equiv & {2 \over c} {1 \over l+1} \int r^l (\rho \vec{v}) \cdot 
(\vec{r} \times \nabla Y_l^{*m}) d^3x \ , \\ 
\end{eqnarray}
where $c$ is the speed of light and $Y_l^m$ are the spherical harmonics.
Since the energy radiated away from the system
is a large fraction of the rotational energy of stars (measured in the
inertial frame), $E_{\rm rot}$, 
the time scale of the gravitational wave emission, $\tau_{\rm GW}$, can be 
roughly estimated as
\begin{equation}
\tau_{\rm GW} \sim \left( \frac{1}{E_{\rm rot}} \frac{dE}{dt} \right)^{-1} \ . 
\end{equation}
This estimation can be applied to all rotating stars irrespective of their
rotation laws. 

As seen from the eigenfunctions of the perturbed velocity fields, oscillations
with large amplitudes for differentially rotating stars are confined to the 
surface region where the density is small. On the other hand, the 
eigenfrequencies of differentially rotating stars are not different 
significantly from those of uniformly rotating stars. Thus the time 
derivatives of the {\it current multipole moments} of differentially rotating 
stars are considerably smaller than those of uniformly rotating stars. 
It implies that the time scale $\tau_{\rm GW}$ for differentially rotating
stars would be longer than that estimated for uniformly rotating stars.
Therefore, the instability  of r-mode oscillations of differentially
rotating stars would not be so significant as that of corresponding
uniformly rotating stars having the same amount of rotational energy.  
Of course, this must be checked by computing the growth rates of
oscillations quantitatively by using such a scheme as was employed to analyze
the f-mode instability of rapidly rotating stars \cite{ye95}.

\subsection{R-mode oscillation of differentially rotating stars and the 
strange nature of the r-mode oscillation of the perturbational approach 
in general relativity}

As discussed in Introduction, r-mode oscillations of neutron stars
should be analyzed in the
framework of general relativity.  Although r-mode oscillations in general 
relativity have been investigated by several authors, there appear some
features which do not exist in {\it uniformly rotating} Newtonian stars 
\cite{k98,kh99,bk99,l99,laf00}.  One of them is related to the strange 
nature of the eigenvalue problem if {\it barotropic} or {\it isentropic}
configurations are treated. Some authors conclude the eigenvalue problem
becomes singular \cite{k98,kh99} and other authors claim that the 
system results in an over-determined state \cite{l99,laf00}.

The strange nature of the eigenvalue problem for barotropic stars in general 
relativity arises from appearance of the following term in the coefficient of 
the highest order derivative term of the master equation if the slow rotation
approximation is employed:
\begin{equation}
{\hat \sigma_{\rm GR}} \equiv \sigma - m \left( \Omega 
- \frac{2 (\Omega-\omega(r,\theta))}{\ell (\ell+1)} \right) ,
\end{equation}
where $\omega(r,\theta)$ is the dragging of the inertial frame
of rotating general relativistic configurations and $\ell$ is the
index of spherical harmonics. 

If we introduce the following {\it effective} angular velocity
\begin{equation}
\Omega_{\rm eff} \equiv \Omega 
- \frac{2 (\Omega-\omega(r,\theta))}{\ell (\ell+1)},
\end{equation}
then we can write the quantity $\hat \sigma_{\rm GR}$ as follows:
\begin{equation}
{\hat \sigma_{\rm GR}} = \sigma - m \Omega_{\rm eff} .
\end{equation}
Here, even if the stellar angular velocity, $\Omega$, is constant value, 
the effective angular velocity, $\Omega_{\rm eff}$ depends on the location. 
Thus the peculiar nature of the eigenvalue problem in general relativistic
slow rotation approximation can be essentially the same phenomena of 
appearance of corotation points for differentially rotating stars which has 
been discussed in the previous section.

In Figure~7, distribution of the quantity $\Omega_{\rm eff}/\Omega$ 
for $\ell = m = 2$  on the equatorial plane is shown for uniformly rotating 
general relativistic polytropes with $N = 1.0$. Relativistic polytropes 
are defined by the following relation:
\begin{equation}
  p = K \varepsilon^{1+1/N},
\end{equation}
where $\varepsilon$ is the energy density. The strength of gravity is measured 
by the ratio of the pressure to the energy density at the center \cite{keh89}.
Models shown in Fig.~7 are configurations with $N=1$ and 
$p_{\rm c}/\epsilon_{\rm c} = 0.5$ 
which corresponds to the typical parameter for neutron stars. 
Here $p_{\rm c}$ and $\varepsilon_{\rm c}$ are the central values
of the pressure and the energy density, respectively.  Two curves correspond 
to equilibrium models with $r_{\rm p} = 0.97$ and $0.59$.  As seen from this 
figure, although rotation speed of two configurations is quite different,
the quantity $\Omega_{\rm eff}/\Omega$ shows little difference.
Consequently, if the eigenfrequencies of the r-mode for general relativistic 
configurations behave similarly as those in Newtonian models, in other words, 
if the eigenfrequencies tend to decrease as the stars rotate rapidly, 
almost the same behavior as those of differentially rotating Newtonian
stars may be expected.

Therefore, there arises a possibility that the quantity 
${\hat \sigma_{\rm GR}}$ can be negative throughout the star 
and does not change the sign for rapidly 
rotating configurations as differentially rotating Newtonian stars do. 
In other words, the ordinary type r-mode oscillations may be possible for 
rapidly rotating general relativistic stars. 

Moreover, as our present numerical results show, if the corotation point is 
about to appear in the star, it becomes very difficult to solve ordinary type 
r-mode oscillations, i.e. r-mode oscillations with discrete eigenfrequencies 
and this seems consistent with the general relativistic analysis under the 
slow rotation approximation. Furthermore, the eigenfunctions for such 
configurations behave like a delta function as seen from Fig.~2.

We may need to explain the reason why some authors \cite{l99,laf00} could 
solve r-mode oscillations for isentropic stars in the post-Newtonian
approximation.  As our numerical results show, weakly or intermediately 
differentially rotating configurations do not suffer from appearance of 
corotation points. Thus for weakly relativistic configurations, there can be
the ordinary type r-mode oscillations, although they do not have exactly 
the same character as that of Newtonian configurations \cite{l99,laf00}.

From our "numerical experimental data" in this paper, there is a possibility 
of obtaining ordinary type r-mode oscillations for {\it rapidly rotating 
general relativistic} configurations, if one will be able to develop a scheme 
to handle linearized equations of the general relativistic fluid equations as 
well as the linearized Einstein equation for rapidly rotating stars.

\subsection{Summary}

We have succeeded in developing a new scheme which can be used to solve 
r-mode oscillations for stars which rotate differentially.
This is a necessary development in order to study the scenario that neutron 
stars have lost their angular momentum via gravitational waves just after 
their birth by the r-mode instability. 

By analyzing the r-mode oscillations of differentially rotating polytropes,
we have shown that the eigenvalue problem is likely to become singular for 
configurations with large degree of differential rotation.


\begin{figure}
       \centering\leavevmode
       \psfig{file=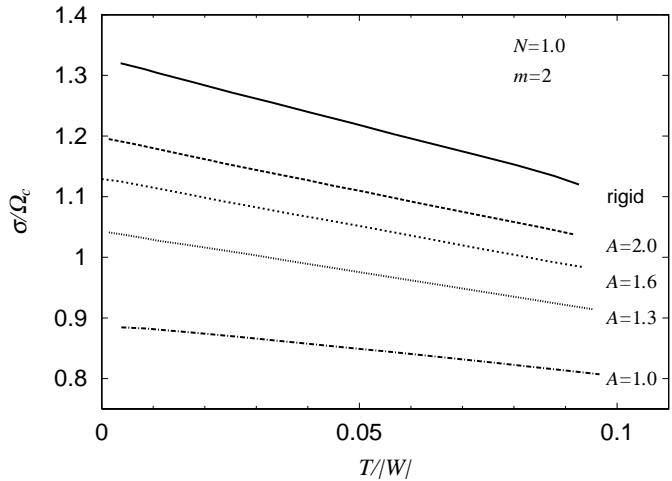,width=9.5cm,angle=0}
\caption{Eigenfrequencies for $m=2$ mode oscillations of $N=1$ polytropic 
equilibrium sequences with different values of $A$ are plotted against the 
ratio of the rotational energy to the absolute value of the gravitational
energy, $T/|W|$.  The attached number to each curve is the value of $A$
along the sequence. The eigenfrequency is normalized by using the
central value of the angular velocity $\Omega_{\rm c}$.
Terminal points of the curves for larger value of $T/|W|$ correspond to final 
models beyond which our numerical code could not give solutions of 
r-mode oscillations.
}
\end{figure}

\begin{figure}
       \centering\leavevmode
       \psfig{file=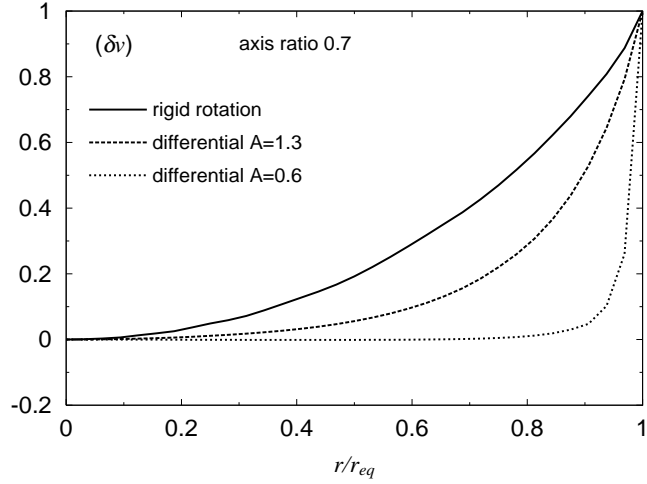,width=9.5cm,angle=0}
\caption{The $\theta$-component of the perturbed velocity on the stellar
equatorial plane is plotted against the distance from the center for $N = 1$ 
polytropes and the $m = 2$ mode oscillation. Three curves show distributions
of the $\theta$-component of the fluid velocity with different degree of
differential rotation: the solid curve for a rigidly rotating configuration,
dashed curve for a configuration with $A = 1.3$ and the dotted curve for a 
configuration with $A = 0.6$. 
The values of the axis ratio for the three models are the same,
$r_{\rm p} = 0.7$, and the values of $T/|W|$ are $T/|W| = 0.073, 0.082$
and $0.080$ for the uniformly rotating model, the model with $A = 1.3$  
and the model with $A = 0.6$, respectively.
It should be noted that the large amplitude oscillations are confined to a 
narrow surface region for the highly differentially rotating model.}
\end{figure}

\begin{figure}
       \centering\leavevmode
       \psfig{file=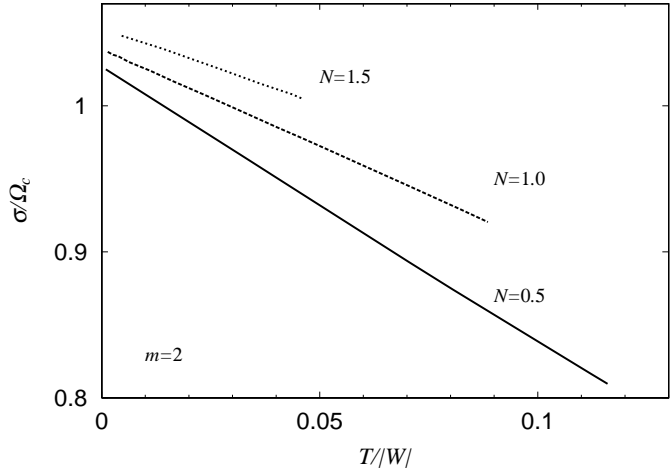,width=9.5cm,angle=0}
\caption{
Same as Figure~1 but for different values of $N$. Attached values to the 
curves are the polytropic indices, $N = 0.5, 1.0$ and $1.5$. 
The rotation law is that of the j-constant law with $A=1.3$.
}
\label{fig:EOS}
\end{figure}

\begin{figure}
       \centering\leavevmode
       \psfig{file=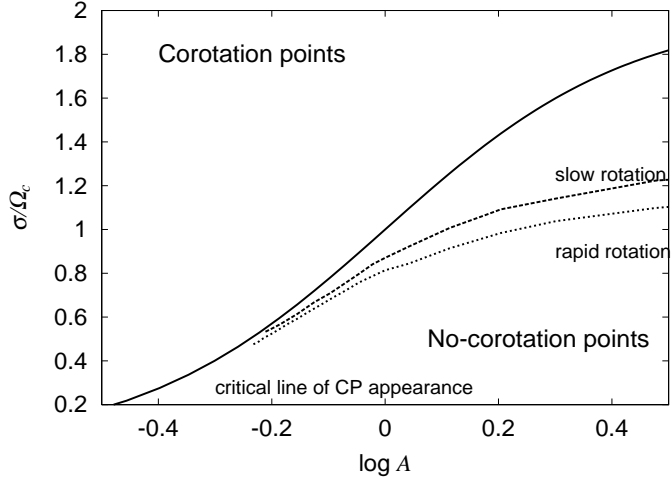,width=9.5cm,angle=0}
\caption{The critical curve for appearance of corotation points
and eigenfrequencies of two equilibrium sequences are plotted
against the value of $\log A$. Solid curve denotes the critical curve.
Dashed and dotted curves show the eigenfrequencies for the equilibrium 
sequences with $r_{\rm p} = 0.95$ ({\it slow rotation}) and $0.70$
({\it rapid rotation}), respectively.
Terminal points of eigenvalue curves correspond to models with 
$A=0.615$ for the $r_{\rm p} = 0.95$ sequence, and $A=0.585$ for
the $r_{\rm p} = 0.70$ sequence, respectively.
The values of $T/|W|$ along the sequences are not exactly but roughly the same.
For $r_{\rm p} = 0.95$ sequence, $T/|W| = 0.023 \sim 0.031$ and
for $r_{\rm p} = 0.70$ sequence, $T/|W| = 0.078 \sim 0.080$.
These terminal points of the curves for smaller values of $A$ correspond to 
final models beyond which our numerical code could not give solutions 
of r-mode oscillations.
}
\label{fig:critforrot1}
\end{figure}

\begin{figure}
       \centering\leavevmode
       \psfig{file=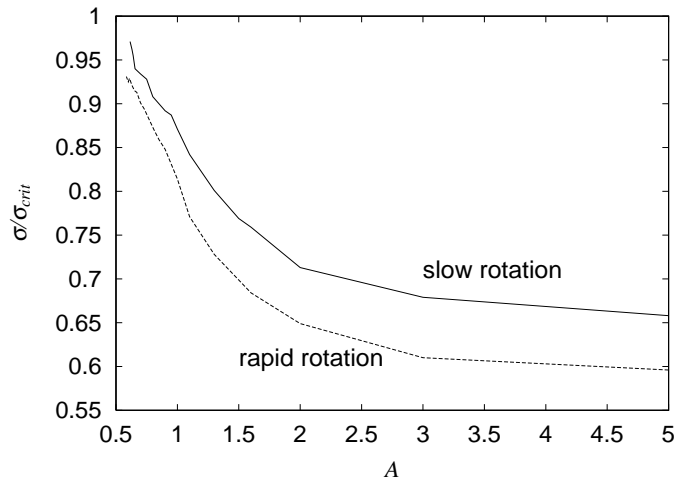,width=9.5cm,angle=0}
\caption{The ratio of the eigenfrequency to the critical value for the
appearance of corotation points is plotted against the value of $A$.
The maximum value of this ratio reaches 97\% of the critical value for 
the appearance of corotation points along the $r_{\rm p} = 0.95$ sequence.}
\label{fig:ratio1}
\end{figure}

\begin{figure}
       \centering\leavevmode
       \psfig{file=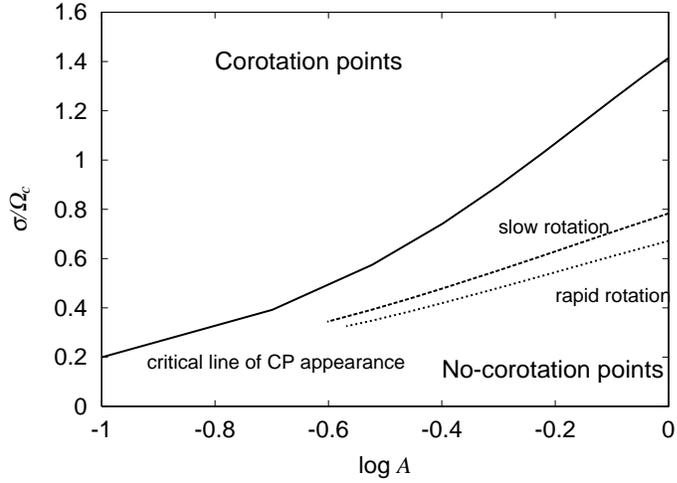,width=9.5cm,angle=0}
\caption{
Same as Fig.~\protect\ref{fig:critforrot1} but for the v-constant
rotation law. 
Dashed and dotted curves show the eigenfrequencies for 
the equilibrium sequences with $r_{\rm p} = 0.95$ and $0.60$, respectively.
The values of $T/|W|$ along the sequences are $T/|W|  = 0.010 \sim 0.011$
for the $r_{\rm p} = 0.95$ sequence and $T/|W| = 0.109 \sim 0.113$ for
the $r_{\rm p} =0.60$ sequence.
}
\label{fig:critforrot2}
\end{figure}

\begin{figure}
       \centering\leavevmode
       \psfig{file=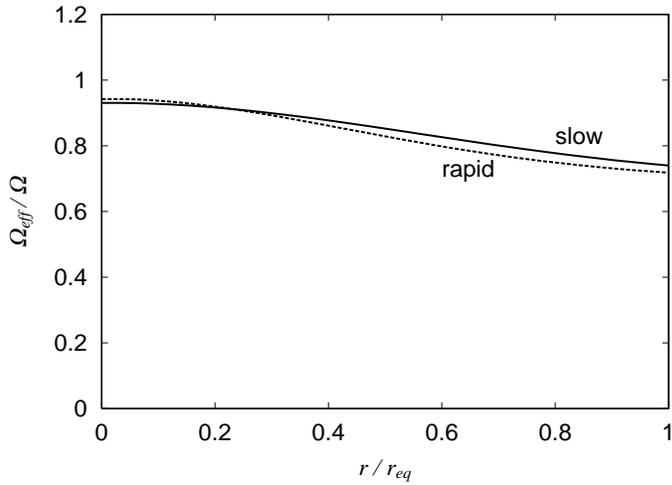,width=9.5cm,angle=0}
\caption{
Distribution of the quantity $\Omega_{\rm eff}/\Omega$ is shown
on the equatorial plane of uniformly rotating general relativistic
polytropes with $N = 1.0$. The strength of gravity parameter is large,
i.e. $p_{\rm c}/\varepsilon_{\rm c} = 0.5$. The solid and dashed
curves show the models with $r_{\rm p} = 0.97$ (slow) and $ 0.59$ (rapid), 
respectively.
}
\end{figure}



\end{document}